\documentclass[preprint,10pt,5p]{elsarticle}
\usepackage{listings}
\usepackage{color}
\usepackage{enumitem}
 \usepackage[tbtags]{amsmath}
 \usepackage{tikz}  
\usepackage{hyperref}
\usepackage{lineno}
\modulolinenumbers[1]
\usepackage{fancyvrb}

\usetikzlibrary{shapes,arrows,calc,positioning}
\definecolor{codegreen}{rgb}{0,0.6,0}
\definecolor{codegray}{rgb}{0.5,0.5,0.5}
\definecolor{codepurple}{rgb}{0.58,0,0.82}
\definecolor{backcolour}{rgb}{0.99,0.99,0.99}
 
\lstdefinestyle{mystyle}{
    backgroundcolor=\color{backcolour},   
    commentstyle=\color{codegreen},
    keywordstyle=\color{magenta},
    numberstyle=\small\color{codegray},
    stringstyle=\color{codepurple},
    basicstyle=\footnotesize,
    breakatwhitespace=false,         
    breaklines=true,                 
    captionpos=b,                    
    keepspaces=true,                 
    numbers=left,                    
    numbersep=5 pt,                  
    showspaces=false,                
    showstringspaces=false,
    showtabs=false,                  
    tabsize=2
}
 
\usetikzlibrary{shapes.geometric, arrows}

\usepackage{graphicx}

\usepackage{amssymb}

\usepackage{lineno}
\modulolinenumbers[1]

\newcounter{bla}

\journal{Computer Physics Communications}

\begin{document}

\begin{frontmatter}



\title{FaVAD: A software workflow for characterisation and visualizing of defects in crystalline structures}


\author[a]{Udo von Toussaint\corref{author}}
\author[a]{F. J. Dom\'inguez-Guti\'errez\corref{author2}}
\author[b]{Michele Compostella}
\author[b]{Markus Rampp}

\cortext[author] {Corresponding author.\\\textit{E-mail address:} udo.v.toussaint@ipp.mpg.de}
\cortext[author2] {Corresponding author.\\\textit{E-mail address:} javier.dominguez@ipp.mpg.de}
\address[a]{Max-Planck-Institut f{\"u}r Plasmaphysik, Boltzmannstrasse 2, 85748 Garching, Germany}
\address[b]{Max-Planck Computing and Data Facility, Giessenbachstrasse 2, 85748 Garching, Germany}

\begin{abstract}
The analysis of defects and defect dynamics 
in crystalline materials is important for fundamental science and for
a wide range of applied engineering. With increasing
system size the analysis of molecular-dynamics simulation data becomes non-trivial. 
Here, we present a workflow for semi-automatic identification and classification of
defects in crystalline structures, combining a new approach for defect description with several
already existing open-source software packages.
%
Our approach addresses the key challenges posed by the often
relatively tiny volume fraction of the modified parts of the sample, thermal motion
and the presence of potentially unforeseen atomic configurations (defect types) 
after irradiation. 
The local environment of any atom is converted into a rotation-invariant 
descriptive vector ('fingerprint'), which can be compared to known
defect types and also yields a distance metric suited for
classification. 
Vectors which cannot be associated to known structures indicate new types of defects.
As proof-of-concept we apply our method 
on an iron sample to analyze the defects caused by a collision cascade induced by a 10~keV
primary-knock-on-atom. 
The obtained results are in good agreement with reported literature values.
%
%
%

\end{abstract}


\end{frontmatter}




{\bf PROGRAM SUMMARY}

\begin{small}
\noindent
{\em Program Title:} Fingerprinting and Visualization Analyzer of Defects (FaVAD). \\
{\em Licensing provisions:} GPLv3. \\
{\em Programming language:} Python 3, Fortran90, and C\texttt{++}. \\
{\em Supplementary material: \href{https://gitlab.mpcdf.mpg.de/NMPP/favad.git}{FaVAD}} \\
{\em Nature of the problem}: The analysis of damage and damage evolution 
in crystalline materials is important for fundamental science and for
a wide range of applied engineering. 
Defects in materials on an atomic level are commonly analyzed by
Wigner-Seitz or topology Voronoi tessellation based methods.
However, these approaches exhibit specific shortcomings, especially at
elevated sample temperatures.
In order to improve upon that, a more robust and quantifiable 
identification and classification approach of known as well as of unpredicted defect structures
is desirable.\\
{\em Solution method:} A fingerprint-like method is proposed to 
analyze in detail the damage in a material augmented with a probabilistic 
interpretation. It is based on the calculation of a descriptor 
vector for each atom in the sample. These vectors represent in a compact form the
individual environments of the atoms.\\
For standard types of defects (i.e. interstitial atoms) 
the corresponding descriptor vectors can be precomputed and used for
rapid classification.
Unexpected or less common defect types can be identified by
applying a principal component analysis to the descriptor vectors.
Vacancies in the material are identified by computing the radii of the
largest empty spheres which can be embedded into the sample, 
followed by a thresholding process.
This new method is easy to use and requires only modest computational 
resources.\\
Finally, the classified defects are
visualized using the open source software VisIt.\\
{\em Features:} The descriptor vectors are computed 
using the command line interface of QUIP with the Gaussian Approximation
potential (GAP) package.
The analysis of the sample is done using Python 
scripts which make extensive use of the numpy package. 
A modified KDTree2 code is employed to calculate the location of single vacancies and voids. 
The program VisIt is used for the visualization of the classified 
point defects in the sample.
We provide a Dockerfile for automatically creating a portable Docker container which
installs all the programs together with a Python script to analyze as an example a
damaged iron molecular dynamics sample.
A shell script to install the programs locally in a Linux-based server or
desktop environment is included also.
\\

\end{small}


\section{Introduction}
\label{sec:introduction}

Over the past decades, there has been a tremendous increase in the size, scale and scope
of atomistic simulations based on molecular dynamics (MD) or density functional theory
(DFT). The increased simulation size challenges many traditional approaches used for the
analysis of the simulation results, e.g. visual inspection, which needs therefore to be augmented by
new tools. This need is especially pronounced in the realm of the interaction of energetic particles
with solids, e.g. ion-solid interaction, plasma-wall interaction or neutronics. 
Due to the fact that the atomistic properties
of these interactions are difficult to assess from experiments only, simulations are indispensable.
However, high particle energy and/or low cross-sections 
often require large simulation sizes for a realistic simulation of the interaction. 
In many cases the formation and evolution of irradiation-induced defects is of special interest
because it modifies atomic arrangements and consequently also affects or determines the physical and chemical properties of the sample \cite{Krasheninnikov,Binder_2004,PhysRevB.9.5008,Ghaly}.
Thus, there is a general need to study defect formation and defect dynamics in
many areas of science, ranging from the optimization of nanomaterials to 
the design of plasma-facing components for fusion devices.

Molecular dynamics simulations are performed to simulate damage in materials 
in a time range from femto- to nanoseconds, and allows to assess the damage on the
atomistic level.
The quantification and classification of the observed defects are of interest 
on its own (e.g. to estimate damage threshold energies) but 
can also be used as input data for other methods, e.g. with longer time 
scaling \cite{C7SC01052D,1402-4896-2011-T145-014036} for multi-scale modelling.
However, as mentioned above in many cases the required sample sizes are very large (e.g. several 
million atoms or more for radiation damage induced by D-T-fusion neutrons) 
and the number of relevant defects is typically in the range
of $\mathcal{O}\left(10-10000\right)$ - thus essentially 
precluding a manual defect identification in many cases (or making it extremely cumbersome and
time consuming).  To address this problem we have developed the FaVAD toolkit which enables a
semi-automated defect analysis also of large crystalline samples. 

Damaged materials are usually analyzed by computing the volume of
Wigner-Seitz cells of the sample atoms
\cite{Atsuyuki,doi:10.1063/1.4849775}. In a subsequent step an association of these volumes
with different defect types (interstitials, Frenkel pairs, etc.) is performed.
A different method based on the geometry of the Voronoi cell faces has also been used for
defect detection in highly damaged materials \cite{LazarE5769,Weinberg}. 
This latter method presents some advantages over the former one due to the additional 
use of information about the undistorted crystal structure like, e.g., body-centered cubic.
However, non-standard defects and their characterization 
present a challenge to the aforementioned methods, especially if atomic displacements
in the crystal due to thermal motion are relevant.

In this paper, we present a new software tool to classify
and quantify material defects 
requiring only modest computer resources \cite{Jav_UvT,domnguezgutirrez2019classification}. 
Our method exhibits several advantages over standard ones: 1) the
calibration of the descriptor-vector computation to consider 
lattice distortions due to thermal fluctuations in the material; 
2) it provides the probability of an atom being a point or other defect
 by describing its particular local atomic environment in the sample;
3) a consistent approach to identify and characterize new defect types;
4) the reliable identification of vacancies and their
representation by calculating their largest empty-sphere volume.
In combination this helps to obtain a better description of sample
damage and damage development. 
The proposed method has been applied to the study of point-defect formation in 
hydrogenated and pristine tungsten samples by molecular 
dynamics simulations. The results are in good agreement with those 
reported in the literature, 
e.g., for the number of created Frenkel pairs
\cite{Jav_UvT,domnguezgutirrez2019classification}.

The article is structured in the following way: In sec. \ref{sec:methods}, we 
briefly describe the numerical approach used to identify material 
defects. It is based on the computation of descriptor vectors and nearest-neighbours distances.
The results obtained are shown in sec. \ref{sec:results}. In sec. \ref{sec:multicomponent} the modifications needed to tackle multi-component systems and 
extrinsic defects (impurities) are outlined.
Finally, concluding remarks are provided in sec. \ref{sec:conclusions}.

\section{Methods and technical procedure}
\label{sec:methods}

\subsection{Descriptor vector calculation}
\label{sec:DV_method}

The descriptor vector (DV) of the $i$-th atom of the material sample, $ \vec{\xi}^{\ i}$,
provides a fingerprint of its local atomic environment. In a first step the non-negative scalar
field $\rho^{{} i}(\vec r)$ of the atom density around atom $i$ is expressed in terms of spherical harmonic functions and a set of basis functions in radial directions. 
To ensure that the representation can be differentiated the densities of each of the surrounding atoms $j$ (essentially delta-peaks at the position of the atoms) are blurred by 
convolution with (truncated) Gaussians. Thus the density is
defined by a sum of a truncated Gaussian density functions with the 
difference vector $\vec r^{\ ij}$
between the atoms $i$ and $j$, entering the exponent \cite{PhysRevB.87.184115},

\begin{eqnarray}
\rho^{{} i}(\vec r) & = & \sum^{\textrm{neigh.}}_{j} \exp 
\left( -\frac{|\vec r-\vec r^{\ ij}|^2}{2 \sigma^2_{\textrm{atom}}} \right) 
f_{\textrm{cut}} \left( |\vec r^{\ ij}| \right)  \label{eq:Eq1} \\
     & = & \sum_{nlm}^{NLM} c^{(i)}_{nlm}g_n(r)Y_{lm}\left(\hat r\right),
\label{eq:Eq2}
\end{eqnarray}
where $\sigma^2_{\textrm{atom}}$ defines the broadening of the 
atomic position (which may depend on the atomic species) 
and $f_{\textrm{cut}} \left( |\vec r^{\ ij}| \right)$ is 
a smooth cutoff function \cite{PhysRevB.90.104108} which removes atoms $j$ with a distance $|\vec r^{\ ij}|$ exceeding $r^{c}_{ij}$ from the density computation around atom $i$.
The expansion coefficients 
are computed using a set of orthonormal radial functions $g_n(r)$
as $c^{(i)}_{nlm} = \langle g_n Y_{lm} | 
\rho^i \rangle$. 
Here $Y_{lm}(\hat r)$ are the spherical harmonics and $\hat r$ is a unit 
vector in the $\vec r$ direction \cite{PhysRevB.87.184115,Jav_UvT}. 
The sum over the order $m$ of the squared modulus of the coefficients $c_{nlm}$
is invariant under rotations around the central atom \cite{PhysRevB.90.104108}. It is given by
\begin{equation}
\vec{\xi}^{\ i} = \left\{ \sum_m
\left(c_{nlm}^i \right)^* c_{n'lm}^i \right\}_{\ n,n',l},
\label{eq:Eq3}
\end{equation}
where $c^{*}_{nlm}$ denotes the complex conjugate of $c_{nlm}$.
Here each component $i$ of the vector $\vec{\xi}$ corresponds to one of the index triplets $\{n,n',l\}$.
The normalized DV $\vec{q}^{\ i} = \vec{\xi}^{\ i}/|\vec{\xi}^{\ i}|$ is used throughout 
this work.
By construction, the DV is invariant to rotation, reflection, translation, and permutation 
of atoms of the same species, but sensitive to small changes in the local atomic 
environment \cite{PhysRevB.87.184115}. 
The DVs are calculated within the multi-body descriptor framework called 'Smooth Overlap 
of Atomic Positions' (SOAP), which implements Eqs. \ref{eq:Eq1} to \ref{eq:Eq3}.
It is part of the QUantum mechanics and Interatomic Potentials (QUIP) package \cite{quip} 
with the Gaussian Approximation potential (GAP) package \cite{PhysRevLett.104.136403}. 
In the supplementary material, we provide a Python script that computes the DVs of a
damaged Fe sample, as an example of our software workflow.

\subsection{Computing the probability of point defects}
\label{subsec:defects}

The difference of two local environments of the $i$-th atom and $j$-th atom 
can be obtained by calculating the distance between the two corresponding 
descriptor vectors, $d = d \left( \vec{q}^{\ i}, \vec{q}^{\ j} \right)$.
Several distance measures are conceivable but we had good success with both, the 
Euclidean distance and the Mahalanobis distance (see below).
Initially, some scale for the distance between atomic configurations is needed.
This scale can be derived in the following way: For a simulation cell with a thermalized
and defect-free sample (i.e. a single crystal) of $N$ atoms the descriptor vectors are computed. 
The mean of the descriptor vectors $\vec{v}\left(T\right)=\frac{1}{N}\sum_{i=1}^{N}\vec{q}^{i}\left(T\right)$ together
with the associated covariance matrix $\Sigma$ of the descriptor-vector components
allow to write the weighted distance as \cite{Maha}
\begin{equation}
    d^M (T) =
    \sqrt{ \left( \vec{q}^{\ i} - \vec{v}\left(T\right) \right)^{\textrm{T}} 
\Sigma^{-1} (T) \left( \vec{q}^{\ i} - \vec{v}\left(T\right) \right)}.
\label{eq:maha}
\end{equation} 

It is worth mentioning that we did not observe a significant decrease in classification performance by 
using only the diagonal elements of the covariance matrix or even an Euclidean distance, 
i.e. by simply replacing the covariance matrix by the identity matrix. These simplifications are also 
beneficial when the inverse of the estimated covariance matrix is ill-conditioned.\\
The definition of the distance difference, $d^M$(T), between 
two local atomic environments leads us to a probabilistic interpretation of the obtained results. Under the simple assumption that $p\left(d^M(T)\right)$ would follow a Gaussian distribution
the likelihood, $p\left(\vec q^{\ i} \mid \vec{v}\right)$, 
of the $i$-th atom displaying $\vec q^{\ i}$ being in a locally undisturbed lattice environment
can be computed using

\begin{equation}
\centering
p\left(\vec q^{\ i} \mid \vec{v}\left(T\right)\right)  = 
P_0 \exp \left[ -\frac{1}{2}d^M(T)^2 \right],
\label{eq:Eq4}
\end{equation}
where $P_0$ is the normalization factor \cite{Jav_UvT}.
However, experience shows that under the present circumstances the distribution of $p\left(\left(d^M(T)\right)^{2}\right)$ follows typically quite well a chi-squared distribution
of degree $K$ \cite{Linden2014}, with $K$ equal to the number of active components of the descriptor vectors. 
Here the corresponding likelihood is then given by
\begin{equation}
\centering
p\left(\vec q^{\ i} \mid \vec{v}\left(T\right)\right)  = 
\frac{\exp \left[ -\frac{1}{2}d^M(T)^2 \right]\left(d^M(T)^2\right)^{\frac{K}{2}-1}}{2^{K/2}\Gamma\left(K/2\right)}.
\label{eq:Eq4a}
\end{equation}
As a consequence, the atoms in a damaged material sample can be 
ranked according to their probability of being in an undisturbed or 
a distorted environment. This allows for a fast screening and identification of 
'interesting' atoms.
Similarly, the distance distribution for any other known defect type can be computed.
Instead of using the atoms of a single crystalline sample as basis for the mean descriptor 
vector $\vec v$ for example a small sample with an interstitial atom 
is thermalized. The sequence of descriptor vectors for this self-interstitial 
atom (SIA) as function of time can then
be used to compute a reference descriptor vector $\vec v_{\mathrm{SIA}}$ for the identification of
interstitial atoms. Examples for several reference descriptor vectors are displayed in Figure \ref{fig:fig3}.

\subsection{Workflow}
\label{subsec:workflow}

\tikzstyle{decision} = [diamond, draw, fill=white!20, 
    text width=4.5em, text  centered, node distance=3cm, inner sep=0pt]
\tikzstyle{block} = [rectangle, draw, fill=gray!20, 
    text width=8em, text centered, rounded corners, minimum height=4em]
\tikzstyle{blockred} = [rectangle, draw, fill=red!20, 
    text width=7em, text centered, rounded corners, minimum height=4em]
\tikzstyle{line} = [draw, -latex']
\tikzstyle{cloud} = [draw, ellipse,fill=gray!20, node distance=3cm,
    minimum height=2em]
\tikzstyle{block2} = [rectangle, draw, fill=blue!20, 
    text width=10em, text justified, rounded corners, minimum height=4pt, minimum width=20pt]

In general, the analysis of the damage in the sample is
done by first identifying sample atoms in undistorted environments and standard point defects 
using a set of precalculated reference descriptor vectors. 
This provides already a classification of 
the vast majority of the atoms in typical cases. 
For the remaining atoms a second analysis step is needed:
The non-standard or unexpected atomic configurations are identified based on
their 'fingerprint' provided by the descriptor vectors which exhibit a too large 
'distance' (i.e. a too low probability $p\left(\vec q \mid \vec{v}\right)$)
to any of the reference descriptor vectors $\vec{v}$. 
If a new type
of defect has been identified its characteristic descriptor vector can then be added to the
set of reference DVs. 
Vacancies and void volumes are identified by the computation of volumes in the sample where the 
distance to the nearest atom exceeds some threshold using a kd-tree approach.
Subsequently the obtained classification of the atoms and the void structure is visualized using standard graphics software. 
In the following sections, we describe the individual steps of the approach in more detail.
For simplicity, we shall use the label 'defect' for atoms which are in an distorted (besides thermal motion) environment. This distortion may either be caused by the atom itself 
(i.e. being an interstitial atom) or by neighbouring atoms (or lack of them), i.e. if the atom is next to a vacant lattice site.  

\subsubsection{Defect identification}
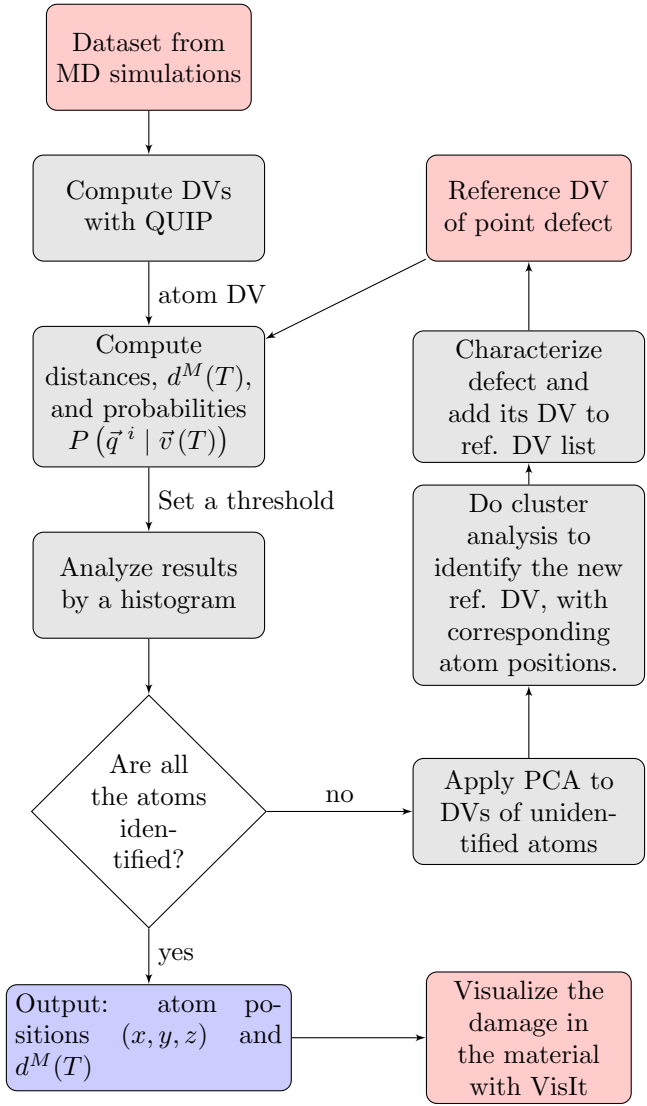
\begin{figure}[!b]
\begin{tikzpicture}[-latex ,node distance = 3cm, auto, on grid]
    \node [blockred] (setup) {Dataset from MD simulations};
    \node [block, below of=setup, node distance=2.0cm] (DV) {Compute DVs with QUIP};
    \node [block, below of=DV, node distance=2.5cm] (distance) {Compute distances, $d^M (T)$, and probabilities $P\left(\vec q^{\ i} \mid \vec{v}\left(T\right)\right) $};
    \node[blockred,right of=DV, node distance=5cm] (refDV) {Reference DV of point defect};
    \node [block, below of=distance, node distance=2.5cm] (hist) {Analyze results by a histogram};
    \node [decision, below of=hist, node distance=3cm] (decide) {Are all the atoms identified?};
    \node [block2, below of=decide, node distance=3cm] (print) {Output: atom positions $(x,y,z)$ and $d^M(T)$};
    \node [block, right of=decide, node distance=5cm] (PCA) {Apply PCA to DVs of unidentified atoms};
    \node [block, right of=distance, node distance=5cm] (add) {Characterize defect and add its DV to ref. DV list};
    \node [block, right of=hist, node distance=5cm] (cluster) {Do cluster analysis to identify the new ref. DV, with 
    corresponding atom positions.};
    \node [blockred, right of=print, node distance=5cm] (visual) {Visualize the damage in the material with VisIt};

    \path [line] (setup) --  (DV);
    \path [line] (DV) -- node {atom DV} (distance);
    \path [line] (refDV) -- (distance);
    \path [line] (distance) -- node {Set a threshold} (hist);
    \path [line] (hist)  -- (decide);
    \path [line] (decide) -- node {no} (PCA);
    \path [line] (decide) -- node {yes} (print);
    \path [line] (PCA) -- (cluster);
    \path [line] (cluster) -- (add);
    \path [line] (add) --  (refDV);
    \path [line] (print) -- (visual);
\end{tikzpicture}
\caption{Workflow for analyzing the damaged material sample by calculating the DV. 
Unclassified atoms are characterized by a principal component analysis (PCA), 
which defines its DV and geometry.}
\label{Fig:DV_wrkflw}
\end{figure}

In order to analyze the damage in a given sample, we follow the workflow 
sketched in Fig. \ref{Fig:DV_wrkflw}. 
The data set (with the information on the atom positions and atom types) 
is taken from the output file (most conveniently provided in the \textit{xyz}-format) 
of the numerical simulation (in our case a molecular dynamics simulation).  
With that information the DVs of all the atoms in the damaged sample (atom DV) are 
computed using QUIP. 
Depending on the total number of atoms, we recommend to use the serial version of QUIP to analyze 
samples of up to $5 \times 10^5$ atoms where the computation of the DV will take only a few 
seconds on a standard desktop computer, 
while the Message Passing Interface (MPI) parallel version of QUIP is suggested for material 
samples with several millions of atoms, taking few minutes on a standard
desktop computer with a multi-core processor.
The next step is to calculate the distances between the atom DVs and the reference descriptor vectors, corresponding to atoms in either an undistorted environment or in/next to known defect structures.
The results can be analyzed, e.g., by using the distribution of the computed distances 
which can be binned into histograms.
This allows to define thresholds for the identification of actual point defects in 
the sample. 
If some atoms are not identified by the initial set of reference DVs, a principal 
component analysis (PCA) \cite{pca_theory} can be applied to the DVs of these atoms. 
Often, the PCA projection yields clustered results already in low-dimensional 
spaces (3- or even 2-dimensional). These clusters are then manually inspected to 
extract the geometry of the underlying defect. 
Its DV is then added to the set of reference DVs for further analyses. 
Once all the atoms are identified, the data is printed out in a \textit{xyz} file with the 
format:  Particle ID, $x$, $y$, $z$, $\vec{d}^M (T)$ .
By following these steps, the output file contains the atomic positions and the distance 
difference to each reference DV of the set of standard defects for each 
atom in the damaged sample. 
The visualization of the sample in 3D can be done e.g. with the software VisIt
by assigning different thresholds to each classification.
In addition, for a fast statistical analysis the visualization software OVITO \cite{Stukowski2009} or similar tools can be useful.

\begin{figure}[!b]
\begin{tikzpicture}[-latex ,node distance = 2.5cm, auto, on grid]
   \node [blockred] (numgrid) {Input parameters};
   \node [block, below of=numgrid] (nn) {Compute NN distance between each sampling point and the atoms};
    \node [blockred, right of=numgrid, node distance=4cm] (setup) {Dataset from MD simulations};
    \node [block,below of=nn] (hist) {Set a lower threshold according to the material properties};
    \node [block,right of=hist, node distance=5cm] (vol) {Output: sampling points denoting the shapes and volumes of the vacancies};

    \path [line] (setup) -- (nn);
    \path [line] (numgrid) -- (nn);
    \path [line] (nn) -- (hist);
    \path [line] (hist) --  (vol);
    
\end{tikzpicture}
\caption{Workflow for calculating vacancy spatial volume by computing nearest neighbours (NN) 
distance between numerical sampling points and the atoms locations.}
\label{Fig:vac_wrkflw}
\end{figure}
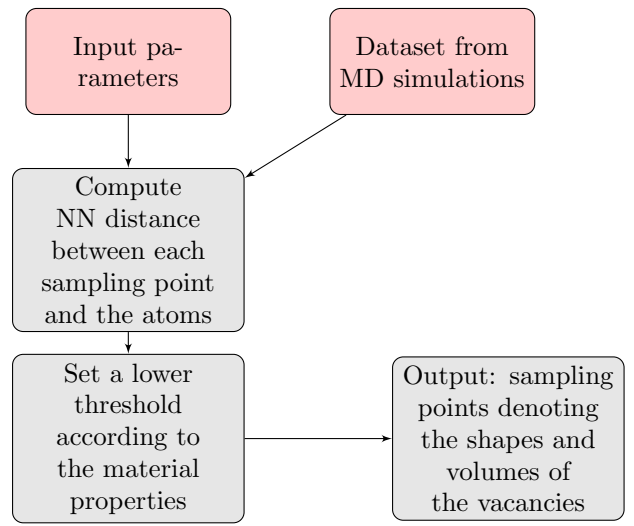

\begin{figure*}[!ht]
\centering
\includegraphics[width=115pt,height=115pt]{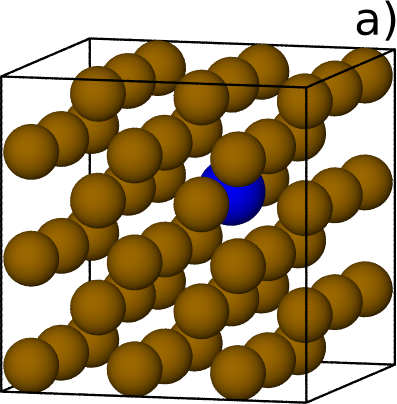} \qquad \qquad 
\includegraphics[width=115pt,height=115pt]{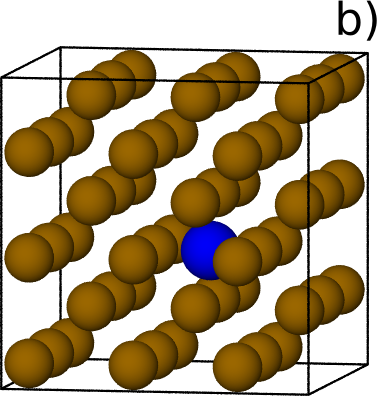} \qquad \qquad 
\includegraphics[width=115pt,height=115pt]{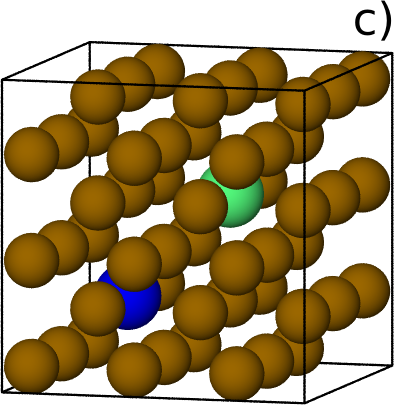}
\includegraphics[width=170pt,height=130pt]{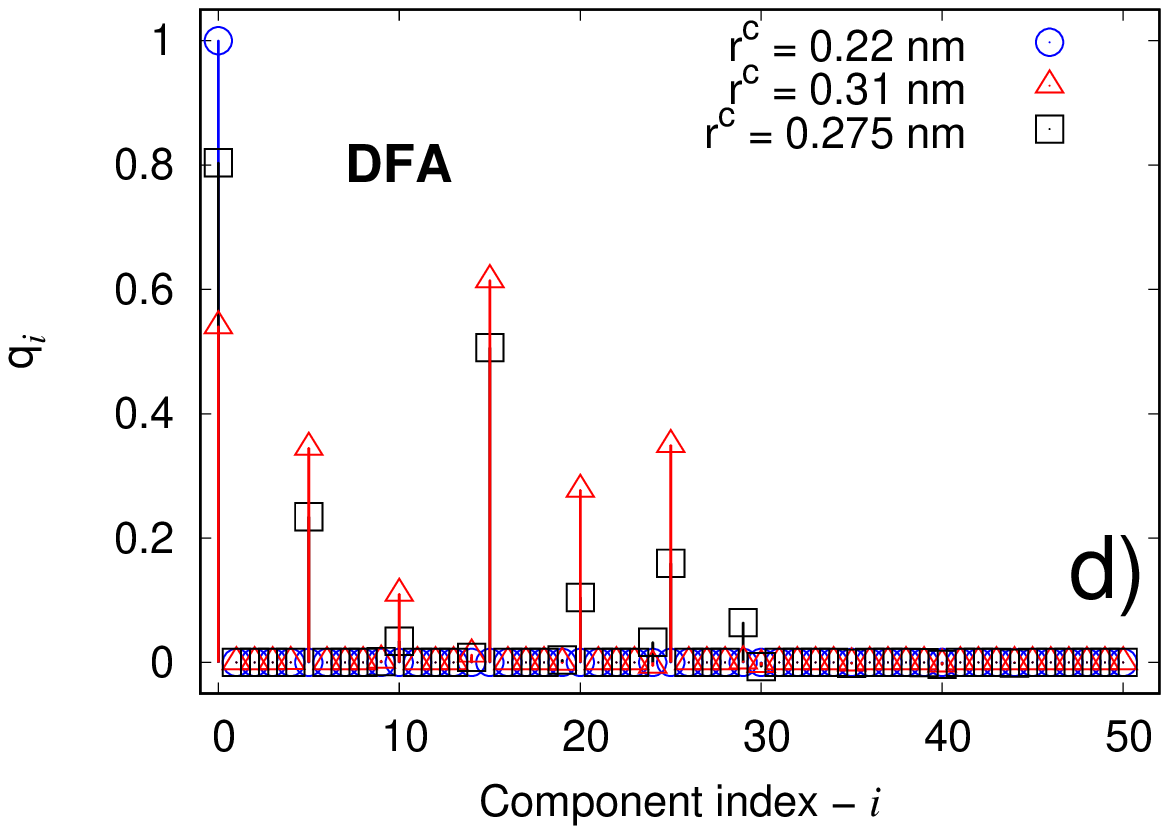}
\includegraphics[width=170pt,height=130pt]{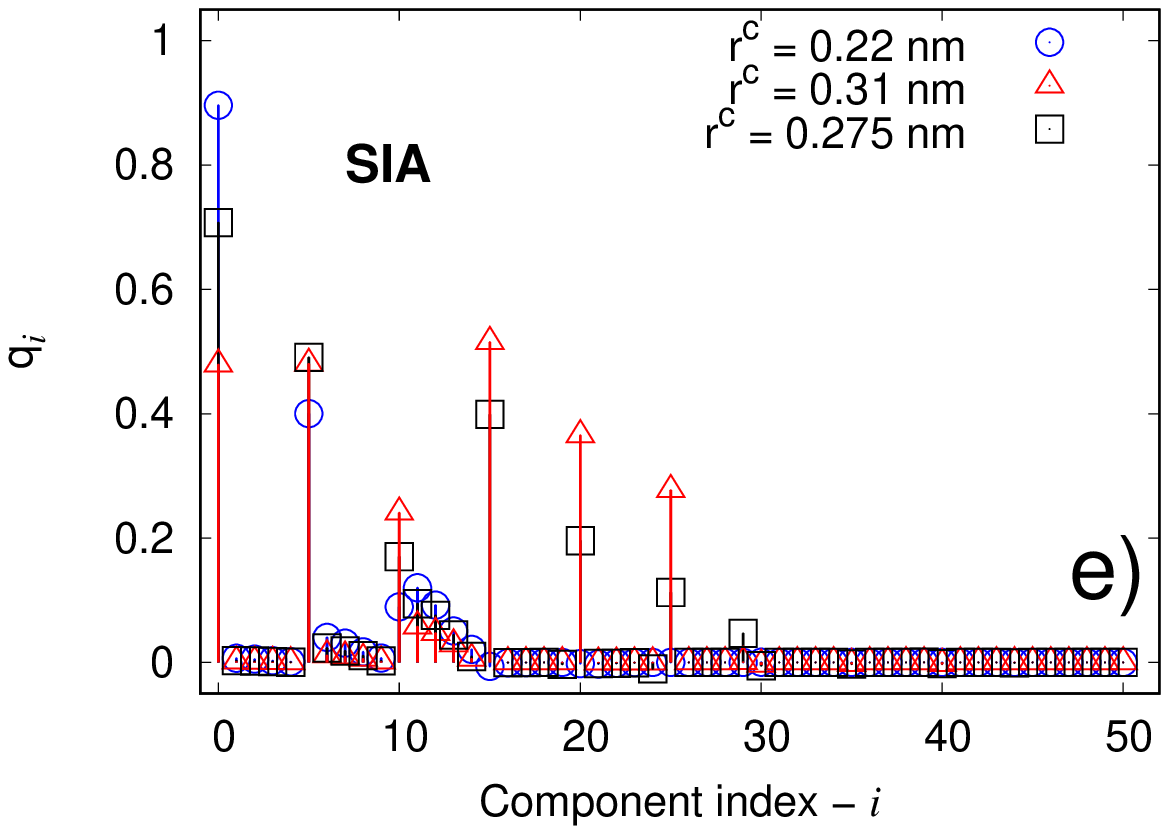}
\includegraphics[width=170pt,height=130pt]{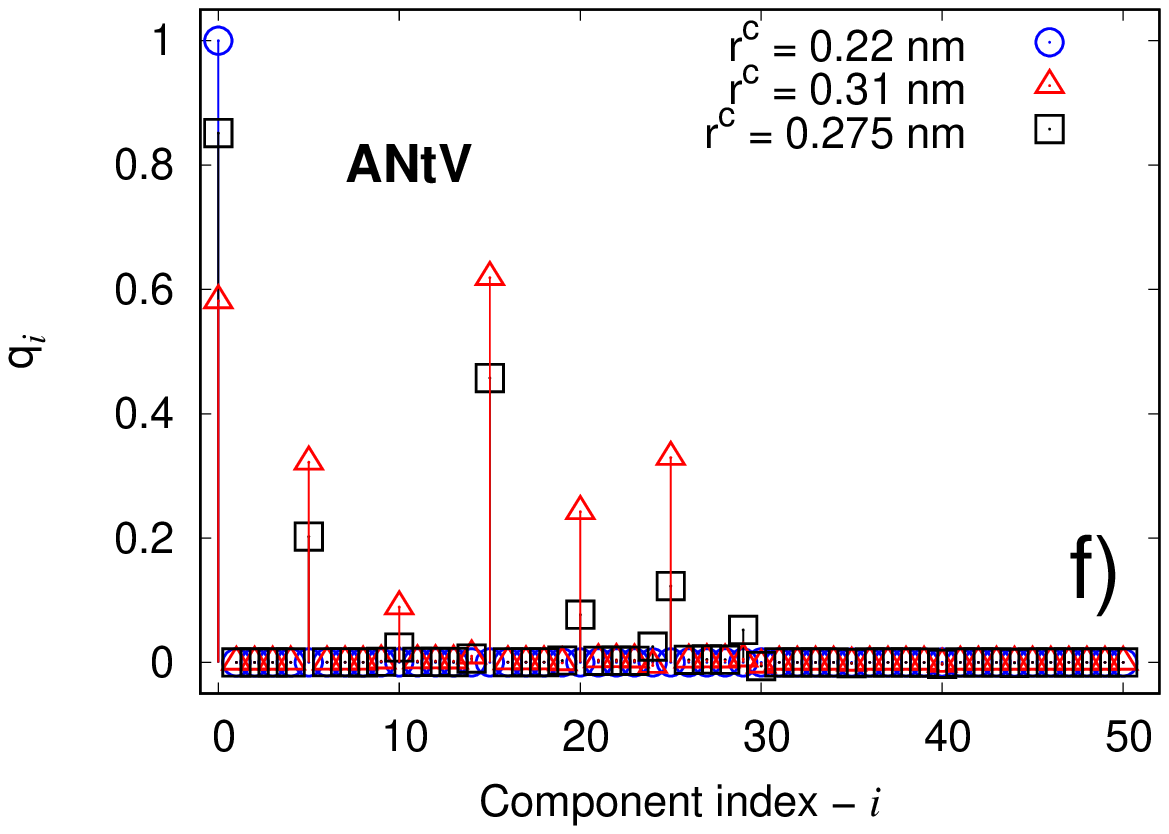}
\caption{Fe geometries used to compute the reference DVs: A Fe atom depicted 
as a blue sphere represents an atom in a lattice position for the DFA classification (panel a), 
in an interstitial site (panel b) for the SIA, and next to a single vacancy (panel c) for the ANtV. 
A green sphere represents a vacancy in the lattice structure.
Amplitudes of reference descriptor vectors as a function of their components are shown in panels d-f) 
for different values of the cutoff parameter $r^{c}$.} 
\label{fig:fig3}
\end{figure*}
\subsubsection{Identification of Vacancies and Voids}
The identification of vacancies with the approach presented in the previous section faces a
challenge: the computation of the descriptor vectors is always anchored at an atom and describes its local environment. However, in a larger void there are no atoms and thus no
information is readily available. 
For that reason a different approach is chosen: 
The identification of vacancies is done by defining a numerical sampling 
grid of $N=N_x\times N_y \times N_z$ points with $N_x,N_y,N_z$ being the number of equispaced points 
in the $x,y$, and $z$ directions, respectively \cite{Jav_UvT}. The 
number of sampling points in each direction is chosen such that the sample-
point spacing (e.g. 0.2~\AA ) is small compared to the
atomic lattice distance. 
Then, a KDTree algorithm \cite{Bentley:1975:MBS:361002.361007} 
is applied to compute efficiently
the nearest-neighbour distance of each of the sampling grid points to 
the closest atom in the sample, followed by a simple thresholding procedure which keeps all the sampling
points where the distance to the nearest atom exceeds the threshold, along the lines of the workflow 
provided in Fig. \ref{Fig:vac_wrkflw}.
We use a modified KDTree ver. 2 code \cite{kennel2004kdtree} for that purpose.
The modified code takes into account the periodicity of the material sample and avoids double counting at the edges of the numerical cell.
Also here, the probability distribution of the nearest atom distances of 
a thermalized defect-free sample can be used to define a suitable threshold (c.f. Fig. \ref{fig:fig6}).

\subsubsection{Visualization}
Once all the point defects are identified and classified and the location and volume 
of the vacancies is computed, the final augmented data can be visualized interactively or 
in batch operation using tools like e.g. VisIt \cite{HPV:VisIt} (c.f. Figure \ref{fig:fig8}).

\section{Results}
\label{sec:results}

As an application of the software workflow, we analyze the effects of a 
simulated neutron impact on an iron bcc-single crystal.
The damage was calculated by performing MD simulations to emulate neutron 
bombardment using a primary knock-on-atom with an initial energy of E=$10$~keV. 
The sample consists of $314928$ Fe atoms and lateral dimensions 
of $15.5 \times 15.5 \times 15.5\,\mathrm{nm}^{3}$. For the analysis only the final atomic 
configuration was known and used.
The data were provided by the Group of K. Nordlund of the University of
Helsinki, details of the sample preparation and molecular dynamics 
simulations are given in Ref. \cite{BYGGMASTAR2018530}.

\begin{figure}[htb]
\centering
\includegraphics[width=0.48\textwidth]{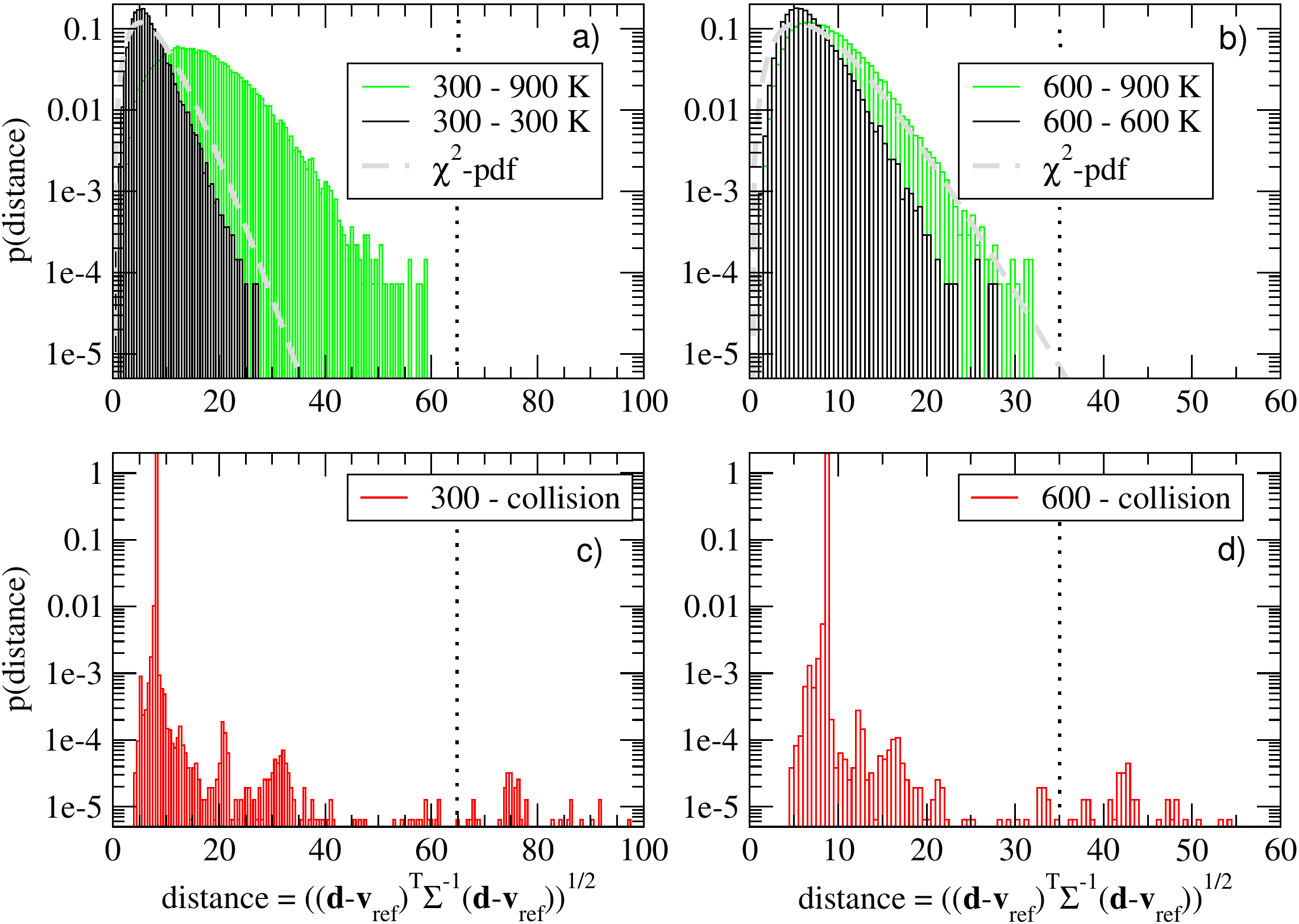}
\caption{(Color on-line) Normalized distance distributions
of the atom DVs with respect to the reference
descriptor vectors for iron lattices at different temperatures. In panel a) the
distributions with respect to a reference DV calculated for a lattice at 300 K is given. Displayed is
the density for the atoms underlying the reference vector calculation, 
i.e. the distance distribution for a
lattice at 300 K. As is indicated by the dashed line, this distribution follows closely the 
expected $\chi^{2}-$density (c.f. Eq.\ref{eq:Eq4a} ). 
Panel a) shows also the distance density of a lattice at 900~K which reflects 
the effect of atomic thermal motion.
The upper boundary of the density provides a good first guess for a reasonable distance threshold 
(highlighted by the vertical dotted line).
In panel b) the same quantities are displayed, 
although this time with respect to the reference descriptor vector computed from a sample at 600~K.
The lower two panels display the distance distribution of the iron sample after the collision cascade. 
There is a clear separation between the large majority of the atoms and 
a small fraction at larger distances. For more details please refer to the text.
}
\label{fig:fig2a}
\end{figure}

In order to define the set of reference DVs, we first consider 
a perfect lattice sample using a small simulation box with 
lateral dimensions of $L_x = L_y = L_z = 3a_0$ with $a_0$ being the lattice 
constant of the material, as shown in Fig. \ref{fig:fig3}a, in our case $a_0 = 0.287$ nm. 
To account for displacement due to thermal motion, reference 
DVs of samples at different temperature ranging from 0~K to about one-third of the
melting temperature of iron (600~K) were computed. 
These form a first set of reference descriptor vectors which are labelled 
as \textbf{DFA} (for Defect-Free-Atom). The obtained distance densities are 
displayed in the two upper panels of Figure \ref{fig:fig2a}.
In the upper left panel the
distributions with respect to a reference DV calculated for an iron lattice at 300 K is given.
Displayed is the density for the atoms underlying the reference 
vector calculation, 
i.e. the distance distribution for a
lattice at 300 K. As is indicated by the dashed line, this distribution follows closely the 
expected $\chi^{2}$-density (c.f. Eq.\ref{eq:Eq4a}). The fit of the $\chi^{2}$-density involves 
only a single free 
parameter $k$, which typically reflects the number of contributing independent components. 
Here the fit yields $k=7$, which is in good agreement with the number of active components in the
descriptor vectors for the present lattice geometry, as can be seen in the panels of Fig.\ref{fig:fig3} (a-c).
In addition, also the distance density of a lattice at 900~K is displayed. 
This roughly corresponds to half of the
melting temperature of iron. At this temperature, the atomic thermal movement is already quite pronounced, although the lattice
structure is still preserved. Thus, the distance distribution provides a good indication for a reasonable choice
of a distance threshold above which atoms can be considered to be in a distorted environment. 
A possible choice
is indicated by the vertical dotted line. In the upper right panel the same quantities are displayed,
but this time with respect to the reference descriptor vector computed from a sample at 600~K.
Please note that the shape and extent of the histograms of the atoms underlying the 
respective reference (300~K in the left panel and 600~K in the right panel) are roughly 
similar (due to the scaling of the descriptor vectors by the variances).
The lower two panels of Figure \ref{fig:fig2a} display the distance distribution of the iron sample 
after the collision cascade. 
A noticeable feature is the pronounced spike at distances of around 10. 
This is caused by the large number of atoms
in almost ideal lattice positions (the sample was initially at 0~K). 
There is also a clear separation between the 
large majority of the atoms and a small fraction (please note the logarithmic ordinate) at larger distances. 
These atoms (about 50 out of 314000) are in or next to a distorted environment. 
Although with that information the number of atoms to be 
looked at in more detail can now be narrowed considerably, 
the nature of the distortion is, at this stage, not yet known.

A second defect classification set labeled as \textbf{SIA} (Self-Interstitial-Atom), 
is computed for a sample with a single Fe atom inserted at an interstitial site into the perfect 
lattice (c.f. Fig. \ref{fig:fig3}b), followed by a
relaxation of the atom positions of the sample. For simplicity 
the size of the simulation box was kept constant.
The DV associated with this inserted atom is used to identify atoms at regular interstitial
positions in the damaged sample.
The third elementary defect type for which initially a reference descriptor is computed are vacancies. 
Here the atoms next to a vacancy are considered. The classification \textbf{ANtV} (Atom-Next-to-a-Vacancy)
is established by removing an Fe atom from the perfect lattice sample (depicted as a 
green sphere in Fig. \ref{fig:fig3}c), again followed by a relaxation of the atom positions.
Note that a single vacancy in a bcc-lattice is surrounded by eight atoms in symmetric
lattice positions, 
one of them is represented by a blue sphere in Fig. \ref{fig:fig3}c) \cite{Jav_UvT}. 
The ANtV-DV can be used to screen samples for atoms neighbouring a vacancy.

This initial set of three reference DVs is used in a first run to analyze the 
damaged Fe sample. Atoms which cannot reliably be assigned to one of the three types 
are then looked at in a second pass.

The three descriptor vectors are provided in 
the supplementary material as:

\begin{enumerate}
    \item \verb+bcc_vector.dat+  for DFA
    \item \verb+interstitial_vector.dat+ for SIA
    \item \verb+type_a_vector.dat+  for the DV of an unclassified defect found by the PCA.
\end{enumerate}


\subsection{Choice of radial cut-off parameter and expansion order}
The cutoff $r^{c}_{ij}$ is an important parameter in the calculation of the DV, and needs 
to be chosen according to the crystal structure of the material sample. A sensible value 
should be larger than the nearest-neighbour distance $r_{NN}$ which for the present case 
of an bcc-iron lattice is given by $r_{NN}=\sqrt{3}\left(0.287/2\right)\,$~nm = 0.2485~nm.

In Figs. \ref{fig:fig3}d-f) we present a comparison of the resulting DVs for 
different values of the cutoff parameter for three different 
atomic environments: DFA, SIA, and ANtV. 
For a cutoff of $r^{c}_{ij} = 0.22$~nm (below the 
nearest neighbour distance) the descriptor vector displays only the radial 
symmetric component (vector component zero) of the 
central atom because no other atom contributes to the atom density for the cases a and c. 
This is different for the case b (interstitial atom). Here the nearest-neighbour atoms 
are closer than the cut-off distance $r^{c}_{ij}$ and thus give rise to 
contributions (c.f. Fig. \ref{fig:fig3}e, component indices above 10) to the descriptor vector.
A value of $r^{c}_{ij} = 0.275$~nm is chosen to describe the atomic environment including 
the first nearest-neighbour atoms (1NN) which are at a distance of around 0.2485~nm.  
To consider also the second nearest neighbour atoms the cut-off distance is 
set to $r^{c}_{ij} = 0.31$~nm, exceeding the Fe lattice constant of 0.287~nm but still well
below the distance to the third-nearest atoms (0.41 nm). Examining the contributing
components of the descriptor vectors it turns out that for the present defect types 
the two different cut-off distances of 0.275~nm and 0.31~nm yield qualitatively similar
descriptor vectors (i.e. concerning the non-zero components) but exhibit differences in
the amplitudes. This can be different for other defect types with lower
symmetry. It can also be seen by comparing the descriptor vectors of the panels d) and e) of
Fig. \ref{fig:fig3}  that interstitial atoms yield a descriptor vector which can easily be 
distinguished from DVs of atoms in undistorted environments (i.e magnitude of 
vector component 20 in comparison with its two neighbours at index 15 and 25). 
In contrast, a comparison of panels d) and f) shows that descriptor vectors of atoms next 
to a vacancy are close (but not identical) to those of DVs of undistorted atoms. 
The reference DV for a Fe atom in lattice position (Fig. \ref{fig:fig3}d) 
and the one for the atom next to a vacancy (Fig. \ref{fig:fig3}f) have the same number 
of non-zero components. 
However, the amplitudes of these DVs are slightly different as well as their covariances.
In the following calculations we have chosen to use DVs computed with a cutoff 
value of $r^{c}_{ij} = 0.275$~nm.

\begin{figure}[!t]
\centering
\includegraphics[width=0.48\textwidth]{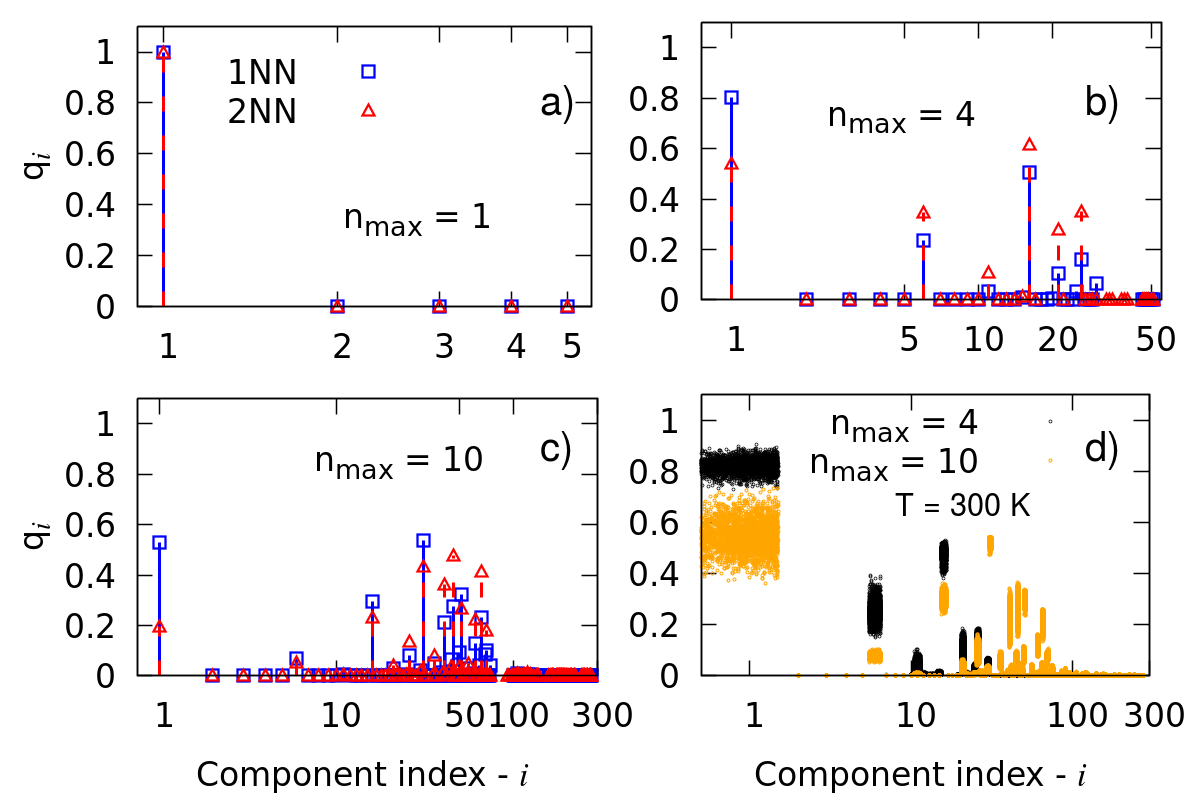}
\caption{(Color on-line) DVs of the defect free atom (DFA) 
classification for different values of
$n_{\textrm{max}}$ and two different cut-off-radii ($r^{c} = 0.275$~nm 
(labelled as 1NN) and  $r^{c} = 0.31$~nm (2NN). 
Beyond a certain $n_{\textrm{max}}$ the relative differences 
between the two representations of the two different density fields
for the most pronounced vector components are decreasing, thus hampering the stability of the
identification. At the same time the number of vector components increases combinatorically 
(please note the logarithmic scale on the abscissa).
} 
\label{fig:fig4}
\end{figure}

Besides the cut-off radius $r^{c}$ also the expansion order $n_{\textrm{max}}$ of the 
spherical harmonics is a free parameter. A too low value yields an expansion which cannot represent 
the density field accurate enough, a too high value becomes increasingly sensitive to small 
and insignificant (thermal) fluctuations. In our experience sensible expansion orders 
are in the range of 4-10. 
In Fig. \ref{fig:fig4}, we present the DVs of the DFA classification at different values 
of $n_{\textrm{max}}$ at cutoff values of 1NN (solid line) and 2NN (dashed line).
The number of components of the DVs increases as a function of 
the $n_{\textrm{max}}$ value because
the number of components is given by the number 
of the ordered pairs $(n,n')$. 
For example, at $n_{\textrm{max}} = 1$ (Fig. \ref{fig:fig4}a) the sequence of components is 
$\{(0,0),(0,1),(0,2),(0,3),(0,4)\}$ for $l_{\textrm{max}}  = 4$. 
At a value of $n_{\textrm{max}} = 4$ (Fig. \ref{fig:fig4}b) and $n_{\textrm{max}} = 10$ 
(Fig. \ref{fig:fig4}c) the DVs exhibit $51$ and $276$
components, respectively. 
Higher order spherical harmonics are increasingly sensitive to thermal fluctuations.
This can be seen in Fig. \ref{fig:fig4}, panel d) where the DV for an atom in a 
defect-free sample at T= 0 K is compared with DVs computed for a 
sample at T= 300 K. Many vector components at higher indices ($j>30$) are
non-zero, but the separation becomes increasingly difficult.
Therefore, we use as default parameters for the present iron 
system $r^{c}_{ij} =0.275$ nm and $n_{\textrm{max}} = 4$.

\subsection{Defect classification and quantification}
\label{subsec:defects_analysis}

\begin{figure}[!t]
\centering
\includegraphics[width=0.48\textwidth,trim={0 15pt  0 0},clip]{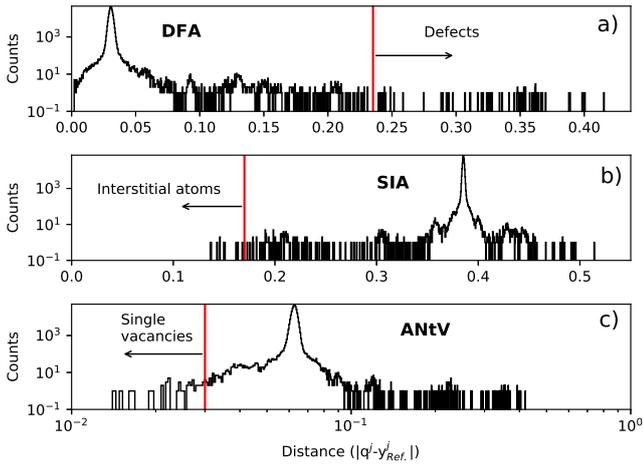}
\caption{Histograms of the distance between 
the DVs of a damaged sample with a reference DV of a lattice 
atom (DFA) in a), in an interstitial site (SIA) in b), and next to a 
vacancy (ANtV) in c).
The red solid lines and arrows indicate the thresholds and distance ranges used for the 
identification of defect atoms, interstitial atoms and atoms next to a vacancy site. 
} 
\label{fig:fig5}
\end{figure}

The DVs of the irradiated Fe sample with $3\times 10^{5}$ atoms can be computed 
within a few minutes usingh QUIP on a standard desktop computer. 
The subsequent step is to identify atoms which are in an distorted environment.
Here we compute the distance distribution of the atom descriptor vectors to the
reference descriptor vector for atoms in undistorted environments using Eq. \ref{eq:maha}.
A typical result is shown in the lower panels of Figure \ref{fig:fig2a}.
The majority of the atoms (please note the log-scale of the ordinate) are in a regular environment.
This is expressed by the dominating peak at small distances. However, the distribution 
exhibits a tail towards larger distances - unlike the distance distribution of the thermalized 
but defect-free sample atoms in the upper panels of Figure \ref{fig:fig2a} - which is thus a
good indication for atoms in or next to defect sites.
The precise distance threshold for an unknown sample has to be determined manually 
by checking above which distance a defect structure is present. 
This threshold estimation typically requires the examination of only a few local environments 
only (using a bisection approach)
and resulted in a distance threshold of 33 for the case of the 600~K DFA-reference 
vector (c.f. panel d of Figure \ref{fig:fig2a}).
It is worth pointing out that even a more simplistic approach works well. This can be beneficial if the
(co-)variances of the descriptor vectors are unknown because insufficient data are at hand. Just setting
the variances to one and using the lattice at 0~K as basis for the DFA-descriptor-vector calculation
yields a distribution as shown in panel a) of Fig. \ref{fig:fig5}. Although the distance scale 
is different
from the one in panel d) of Figure \ref{fig:fig2a}, the same atoms are outlying, i.e. can be 
identified as defects. The derived distance threshold for atoms to be considered as 'non-regular' 
based on the distance distribution provided in Fig. \ref{fig:fig5}~a) is $d_{DFA}^{th}=0.23$ - thus 
yielding a total of 62 atoms to be classified as 'defects' of yet unknown character. 
To determine the type of distortion or defect the atom descriptor vectors are subsequently compared to
the reference vectors of the different defect types.
For example, to decide if an atom is at an interstitial position, the distance of the 
atom DV to the SIA-reference descriptor vector is computed. Atoms for which the distance is 
below some suitable threshold are counted as interstitial atoms. 
In panel b) of Fig. \ref{fig:fig5} the distance distribution of the atom DVs from 
the SIA-reference DV is displayed. Also here manual inspection allows to set a threshold in
a straightforward manner, resulting in $d_{SIA}^{th}=0.17$ and the identification of 15 
atoms as being
self-interstitials.
A similar analysis for the defect type 'atom next to a vacancy' (ANtV) 
classifies 46 atoms as ANtV for a threshold of $d_{ANtV}^{th}=0.03$.
However, unlike the situation for the classification of DFA and SIA, 
the distance histogram of $d_{ANtV}$ does not indicate an evident gap supporting the choice
of the threshold. Closer inspection reveals as underlying reason 
the close similarity of the reference descriptor vectors
for the DFA and ANtV-environments 
for the present bcc-lattice geometry and the selected cut-off-radius. 
This results in a low discriminative power between the few ANtVs and the large number of atoms in a
defect-free position. A potential remedy would be the choice of
a larger cut-off-radius specifically for the identification of ANtV-situations. Here, however, 
we resort to a complementary approach which is focused directly on the identification of empty sites.

\begin{figure*}[!ht]
\centering
\includegraphics[width=150pt,height=130pt]{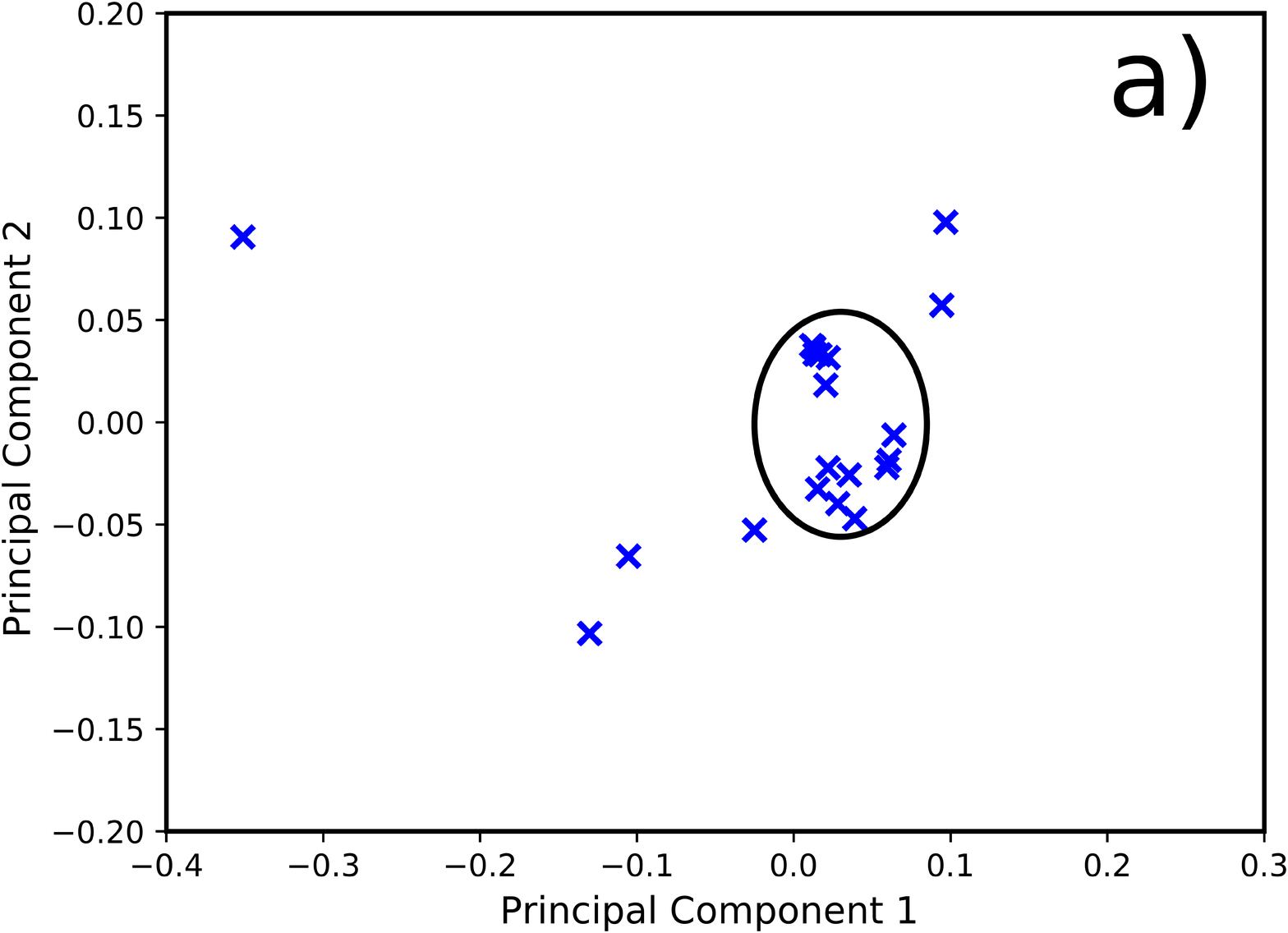}
\includegraphics[width=175pt,height=125pt]{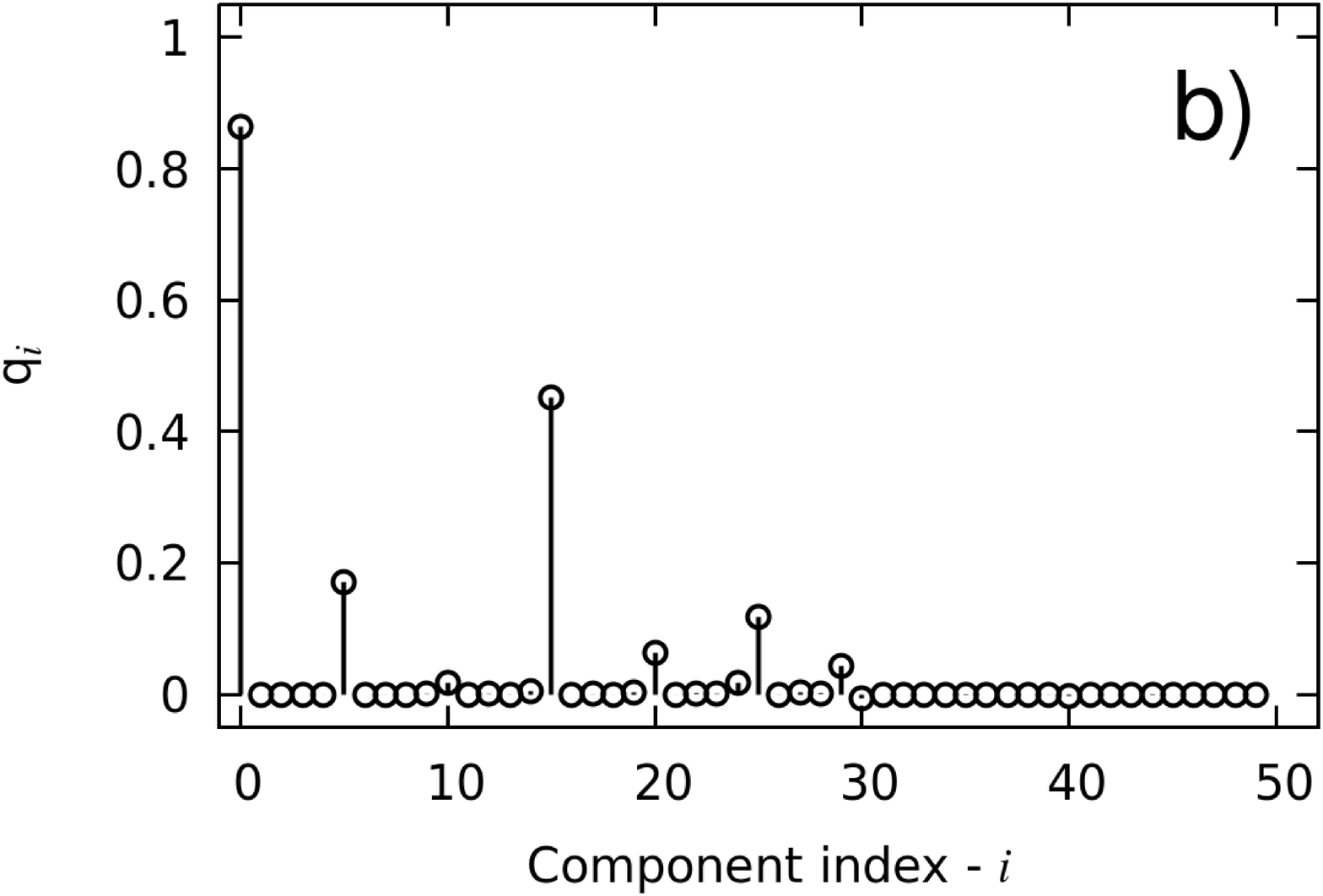}
\includegraphics[width=130pt,height=130pt]{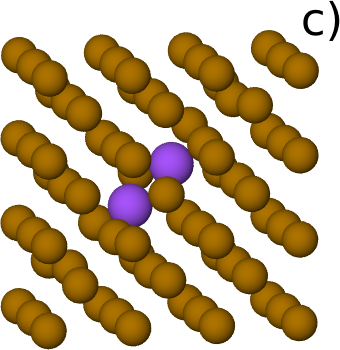}
\caption{(Color on-line) A principal component analysis is applied to the DVs 
of the unclassified atoms and the obtained results are shown in a). 
The DV of the unclassified atom is presented in b), which is found 
by performing a cluster analysis in a). 
The geometry of the unclassified atom is shown in panel c).
Atoms in the new classification are represented by purple spheres, while lattice atoms 
are depicted as brown spheres.
} 
\label{fig:fig7}
\end{figure*}

\subsection{Identification of vacancies and void regions}
\begin{figure}[!b]
\centering
\includegraphics[width=0.48\textwidth]{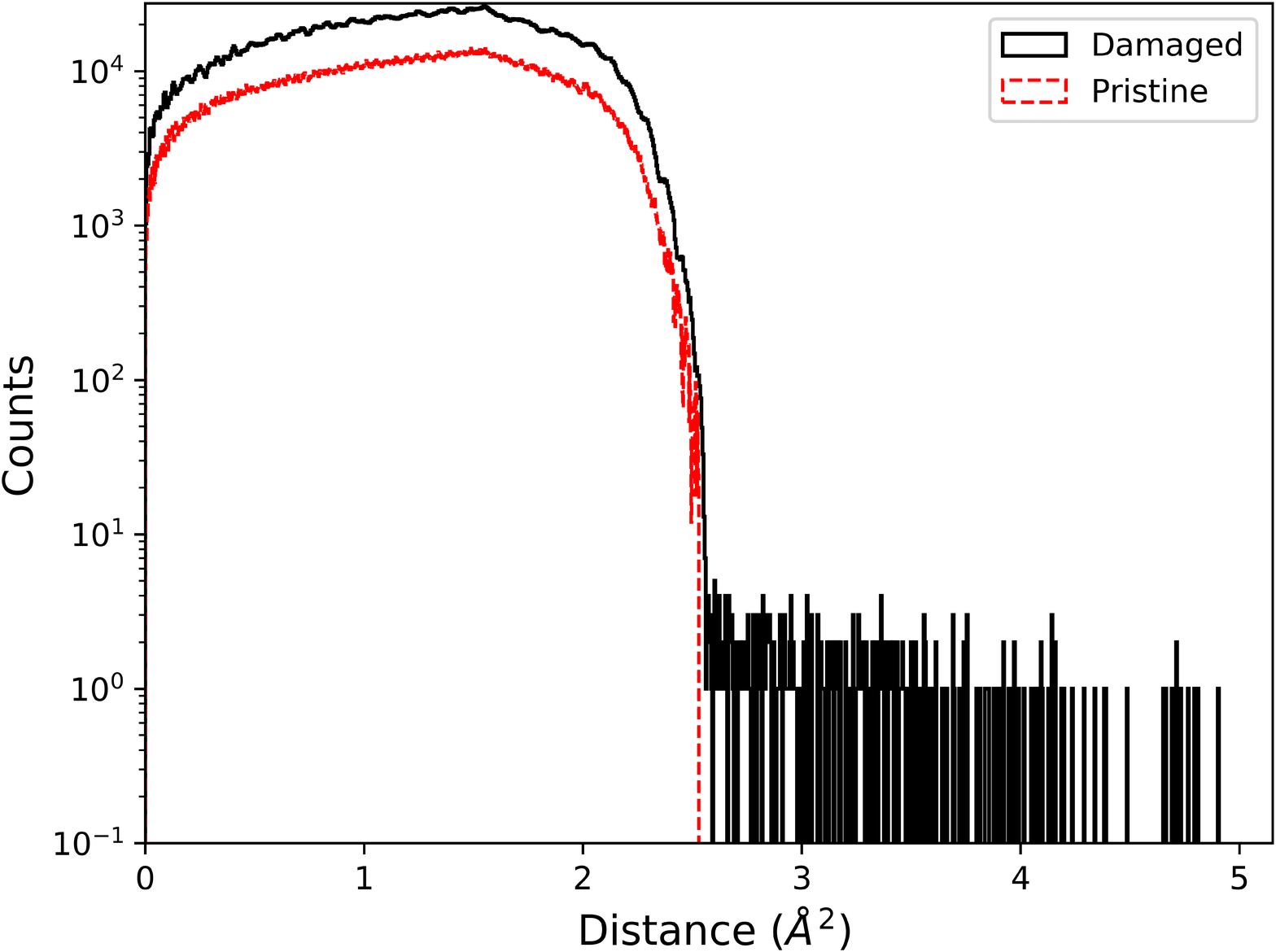}
\caption{Comparison of the nearest distance distribution for a defect-free (pristine) 
sample and the damaged iron sample. The sampling grid had a lattice spacing of 0.0775~nm in
each direction, corresponding to $N=200^{3}$ sampling points. 
For each sample point the distance to the nearest atom is computed using a KD-tree algorithm 
which reduces the numerical effort
from $N^{2}$ to $N\log N$. The sharp cut-off of the distribution of the pristine sample at around 
2.5~\AA${}^{2}$ supports a threshold value slightly above, e.g. 3~\AA${}^{2}$. 
Sample points with distances
to the nearest atom larger than this threshold are considered to be 
located at/near a vacancy site or in a void structure.
} 
\label{fig:fig6}
\end{figure}
The identification of vacancies follows the workflow provided in 
Fig. \ref{Fig:vac_wrkflw}. 
In order to quantify the number of vacancies in the damaged Fe sample, 
we first define a sampling grid of $N = N_{x}\times N_{y}\times N_{z}$ points 
for the simulation box of the damaged Fe sample.
Then the KDTree ver. 2 code \cite{kdtree2} is applied to calculate the nearest neighbour 
distance distribution between the points of the sampling grid and its nearest atom. 
In Fig. \ref{fig:fig6} we show a histogram based on a 
sampling grid with 200 points in each dimension. 
From the result we notice the main part of the distance distribution to be in the 
range of 0 to 2.75 \AA{}$^2$, which corresponds to the (squared) inter-nuclear distance 
to the first nearest-neighbour atoms in the bcc-unit cell. 
Hence, points with a squared distance larger 
than $r^{2}_{th}=3$~\AA{}$^2$ can be 
related safely to the spatial location of the vacancies in the damaged material.
In addition, we note that the numerical grid with 200 points (corresponding to a 
linear spacing of 0.0775~nm) provides enough resolution to 
deduce the spatial extent of an identified void with good resolution.
The computation of the nearest-neighbour distances takes a few minutes for the 
sampling grid with $N=8$~million points on a regular desktop computer.

A straightforward analysis of the set of points exceeding the nearest 
atom distance of $r^{2}_{th}$ yields the number and the location of the vacancies in an 
iterative manner: After identification of the sample point $S_{max}$ with the largest 
nearest atom distance in the set the coordinates of this point are a) assigned as vacancy
position and b) all sample points (including $S_{max}$) with a squared distance smaller than
$r^{2}_{th}$ to $S_{max}$ are removed from the set. Then this procedure is repeated until 
no points are left. Although pretty simple this approach worked quite well in the 
cases considered. However, especially for situations with 
potentially larger void structures also more sophisticated approaches 
like density estimation \cite{Linden2014} or k-means clustering \cite{Bishop2006} can be
used to extract vacancy or void information from the sampling point cloud.
In the present case the provided software identified 25 isolated vacancies and there 
was no indication of larger void structures. Based on the lattice structure and the nearest neighbour distances $25\times 8 \approx 200$ atoms (a few atoms are simultaneously part of another defect) 
can be eventually identified as being next to a vacancy. 
These atoms encompass all 46 atoms which have been identified based on the ANtV-descriptor vector.


\subsection{Characterization of non-standard point defects}
\label{subsec:new_def}

As described above, the analysis of the damage in the sample is performed along the lines of
\begin{itemize} 
\item first, the identification of all the atoms that are in an ordered lattice environment, 
\item second, the identification of atoms which can be associated with a set of standard point defects,
(i.e. interstitial atoms at expected interstitial positions, 
e.g., at tetrahedral or octahedral sites). 
\item subsequently, the localization of vacancies or voids and the neighbouring atoms atoms.
\end{itemize}
However, typically some atoms in a distorted region cannot be associated 
to the initial set of standard defects.
In order to characterize the geometry and DV of these defects in the sample, 
we follow the workflow presented in Fig. \ref{Fig:DV_wrkflw}. 
A principal component analysis \cite{pca_theory} (PCA) is applied to the DVs 
of these special atoms \cite{Alter10101}, projecting the high dimensional descriptor vector in 
a lower dimensional space, typically 2- or 3-dimensional. 
In Fig. \ref{fig:fig7}a) we present the results of this PCA. As can be seen, 
the 2-dimensional projection (and also 3-dim projection, not shown) of the DV form a cluster. 
The mean atomic environment and the mean descriptor vector of the atom descriptor vectors 
yielding the accumulation point 
located at PC1 = 0.01 and PC2 = 0.025 are used to define a new defect structure, 
preliminarily labelled as type-a defect. 
The descriptor vector assigned to the type-a defect is shown in Fig. \ref{fig:fig7}b) and has 
subsequently been added to the set of reference DVs.
In Fig. \ref{fig:fig7}c), we display the geometry of this defect identified by the 
descriptor-vector approach.  In the present case the type-a defect turned out to be an
already known defect type:
the identified geometric arrangement of the displaced atoms is
known as dumbbell, where two atoms (here represented by blue spheres) share a lattice site.
This results also in some distortion of the locations of the surrounding atoms.

Using this new reference descriptor vector to reassess the sample, in total 16 atoms 
are identified as dumbbells. The remaining 6 atoms which have not been classified as a specific defect
type are located close to each other and form a kind of defect cluster. Since there is only a single
occurrence in the sample no further description of this arrangement has been considered worthwhile.
In conclusion the analysis of the iron sample of $3.1\times 10^{5}$ atoms 
yields 15 self-interstitial atoms, 16 atoms in dumbbell position, 25 vacancies and 
a small defect cluster. The present number of Frenkel pairs (i.e. 25) is in good agreement with other 
MD simulations \cite{Nord_NC,STOLLER200022} which reported for similar conditions the formation of 
around 30 Frenkel pairs.


\subsection{Visualization of the damage in the material}
\label{subsec:visualization}

Once the classification of the Fe atoms in the damaged sample is complete, it is possible 
to represent the classified point defects and the location of the identified vacancies 
using the visualization software VisIt \cite{HPV:VisIt}. 
VisIt is a powerful open-source visualization tool for visualizing two- and 
three-dimensional data stored in structured or unstructured meshes. 
It can be used interactively via a Graphical User Interface or by employing a scripting interface 
in the Python (or Java) programming language. The latter simplifies the non-interactive creation of images 
used as frames in a movie.
When plotting the atoms of the Fe sample, we associate a combination of distance differences $d^M$ 
to a color channel in the resulting image, so that each Fe atom is depicted with a color that depends
on its classification. 
The mapping in RGB color space $\left[ 0,255\right]^{3}$, is done as follows:
\begin{flalign*}
  & \textrm{R} : f(- \textrm{$d^M$[Type-a]}; 100, 255) \\
  & \textrm{G} : f(\textrm{$d^M$[ANtV]} - \left[ 1-\textrm{$d^M$[SIA]} \right]; 100, 255) \\
  & \textrm{B} : f(\exp( - \textrm{$d^M$[ANtV]} ); 0, 255),
\end{flalign*}

where the mapping f(x; a,b) is such that it maps the range of x by linear scaling into the interval
$\left[ a,b\right]$.

The final image, composed by the superimposition of the individual color channels,
encodes information on all the reference DVs. 
As an example, a Fe atom is portrayed as a red sphere if characterized by a large value 
of $d^M$[Type-a], which corresponds to the DV found by PCA, and relatively small values for 
$d^M$[SIA] and $d^M$[ANtV]. 
On the other hand, for large values of the $d^M$[ANtV] and small values of $d^M$[Type-a] 
and $d^M$[SIA], the color of the sphere representing the atom is close to green.
Moreover, we scale the size of each sphere according to the cubic root of the volume available to 
the corresponding atom, as calculated by the \verb Voro++ {} code \cite{voropp}, 
which computes the Voronoi cell for each atom individually in the material sample. 

\begin{figure}[!t]
\centering
\includegraphics[width=0.48\textwidth]{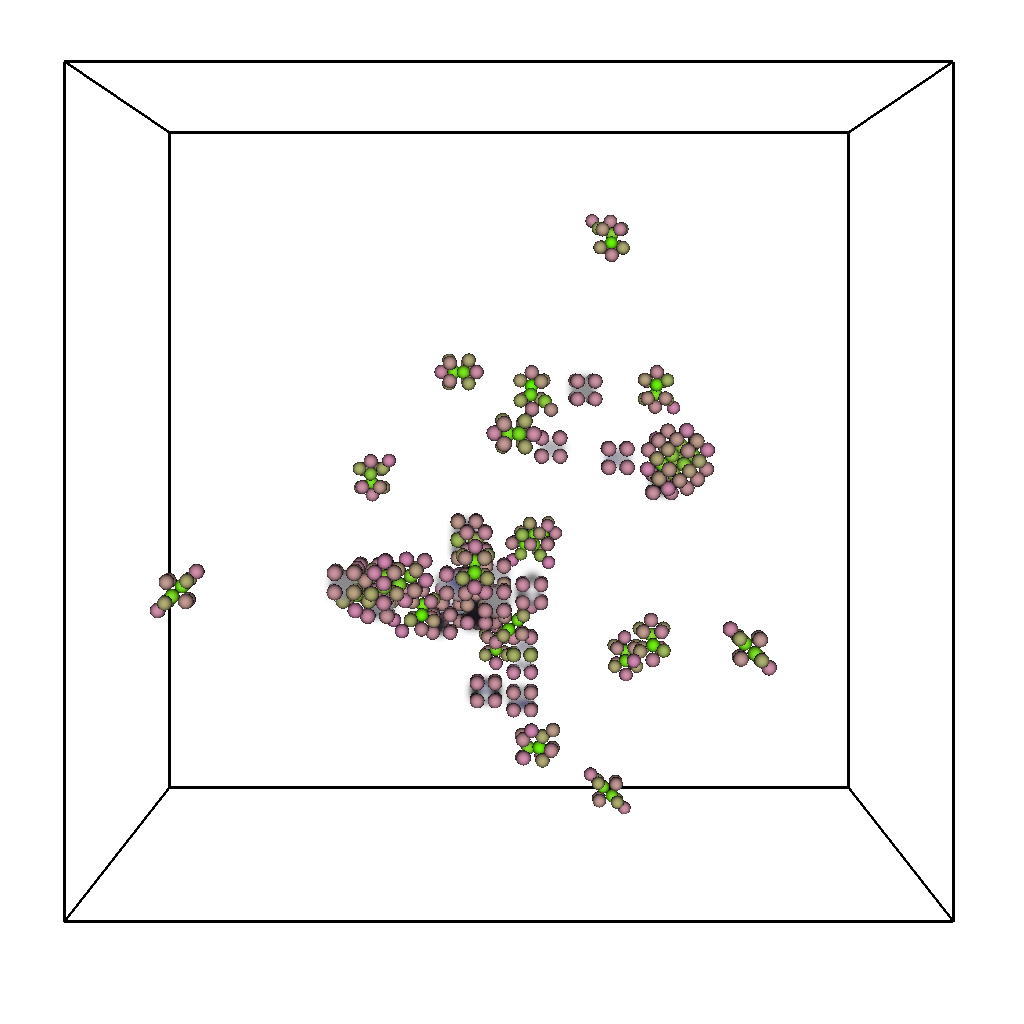}
\caption{Visualization of the damage in the material. Spheres correspond to Fe atoms, 
where each colour channel describes a different combination of $d^M$ for each defect 
classification. 
Cloud-like regions represent vacancies in the volume.} 
\label{fig:fig8}
\end{figure}

We can visualize the location of the vacancies
using a volume plot of the output file of the modified KDTree code which contains the sample lattice
points and their distance to the nearest atom. The points are color coded with distance. 
Points with a distance larger than the threshold distance (here 3 \AA${}^2$) are not displayed,
points with the threshold distance are given in purple and then the purple is changed towards black
up to the sample point(s) with the largest distance.
In order to restrict the number of atoms plotted in the images, we applied an additional threshold:
only Fe atoms with $d^M\textrm{[DFA]}>0.1$ are shown. This renders the atoms in ordered lattice positions invisible.
The resulting image for our reference Fe sample is depicted in Fig. \ref{fig:fig8}. In this visualization
now also the presence of crowdions becomes 
obvious (c.f. the diagonal oriented defect-structure at the very left and at the lower right). 
These have been observed before in the simulation of collision cascades in iron \cite{Stoller2012}.
If appropriate then also this defect type may be added to the set of reference descriptor vectors.

In some instances it may be beneficial to use a slightly reduced distance threshold for visualization (e.g. for the present case 0.05 instead of 0.1). 
Especially at lower sample temperatures this allows to recognize
(and visualize) distorted regions - i.e. atoms which are only slightly displaced by (local) stress fields - within the same workflow.

\subsection{Software Implementation}
\label{sub:soft_impl}

We provide a Dockerfile in the supplementary material which automatically handles the installation 
of QUIP, GAP, and VisIt, and the compilation of the KDTree and Voro++ codes. 
The \verb readme {} file contains the detailed information about the installation of the programs, codes, 
and the use of the docker container that is utilize by our software workflow.
As an alternative to Docker, a traditional bash installer script is included which helps to automatically 
install the required software locally in a standard Linux desktop or server environment.
In addition, a Python script called \verb FaVAD.py {} is included to analyze the material damage 
of a Fe sample as an example. 
This main script needs to be adapted to analyze the damage of a given material.
The following input file and information about the Fe damaged sample are used as an application of our software:

\begin{enumerate}
\item Our software needs a main \textit{xyz} file with all the information about the given damaged sample like 
      atom positions and types, and lattice vectors. 
      This file is read by QUIP to compute the DVs of each atom. 
      In our example this file is called \verb fe_sample.xyz {}, which contains the spatial location of each 
      Fe atom in the damaged sample and the magnitude of the lattice vectors.
\item The set of reference DVs files for standard point defects with the amplitude values 
      for the DV components that are read as input files by our software. 
      These files have to be computed in advance for a particular material, as explained previously and 
      in Ref. \cite{Jav_UvT}. 
      We include the following files in the supplementary material: 
      \verb bcc_vector.dat {}, \\
      \verb interstitial_vector.dat {}, \verb vacancy_vector.dat {}, \\
      and \verb typea_vector.dat {}. These files define the set of reference DVs for analyzing the 
      damage in a Fe sample.
\end{enumerate}

The following generic Python scripts need to be adapted to the atomic constituents. 
In our example, the input parameters are set to the Fe sample properties.
\begin{enumerate}
\item The \verb FaVAD.py {} script is the main script to do the analysis of the damage in the material. 
      It has the information of the sample parameters for computing the DVs with QUIP like material 
      components, the lattice constant, atomic number of the material species, $n_{\textrm{max}}$ and 
      $l_{\textrm{max}}$ values for the spherical harmonics calculations, and the periodicity used in 
      the numerical simulations. 
      It also contains the command lines to obtain the distance difference between the DV of damaged 
      material to those of the standard point defects. 
      
\item All the input files are specified in the \verb parameters.txt {} located in the sample directory.
      
\item The main Python script also includes the input parameters to compute the atoms size with the 
      \verb Voro++ {} code and the points that defines the spatial location of each vacancy with the 
      KDTree code.

\item The VisIt visualization script \verb vis.py {} is included as an input file, which is executed 
      by \verb FaVAD.py {}. 
      It adapts the color code discussed in Sec. 3.3. and generates a video to visualize the damage of the 
      sample at different viewing angles.
\end{enumerate}

The obtained results can be analyzed by histograms and a visualization software.
We also provide the following python scripts in the supplementary material: 
 \verb pca.py {} which applies the PCA for characterizing a new type of defect. 
 In our case, we identified and characterized a 'new' defect-configuration (which turned out to be 
 a dumbbell-arrangement) for the present Fe sample.


\section{Multicomponent systems}
\label{sec:multicomponent}

In many cases of practical relevance the system of interest consists of more than
one element. With some slight modifications the presented approach can 
be applied for the defect analysis in multi-component materials also \cite{Jav_UvT}. 
Naturally, the number of required reference descriptor vectors increases with the
number of possible basic defects. For a binary system with components A and B 
we need to consider - besides the initial ordered lattice - the possible interstitial positions for A and B, 
point defects of both types, etc. Although this is quite manageable for
binary and potentially ternary compounds it may become cumbersome for a larger number
of constituents. We include a python script to compute the DV of a 
binary-component sample in the supplementary material. Some care needs to be applied to
choose the appropriate cut-off radius for the individual descriptor vectors if the
constituents have very different interatomic distances, as e.g. in the case of tungsten and hydrogen \cite{Jav_UvT}.

\section{Concluding remarks}
\label{sec:conclusions}

We have introduced an effective approach for the automated defect analysis 
of crystalline materials. The provided open source toolkit FaVAD will 
back on-the-fly tracing of defect formation as well as spatial and temporal
defect evolution in large scale molecular dynamics simulations or DFT-studies. 
Besides the atom positions the
toolkit requires in its standard setting only inputs which are either readily available 
(like lattice constants) or straightforward to compute (like equilibrium reference structures).
This makes FaVAD attractive for all users interested in the analysis and visualization of defects.
Our results demonstrate that the framework can not only be employed to detect known defect structures,
but can also be used to identify and describe new defect types using a principal component analysis
of the atom descriptor vectors.
In addition, the descriptor-vector approach 
allows to assign a standardized (reference) fingerprint to known or new defect structures, thus 
providing the capability to set-up a database for defect characterisation.

We hope that the openly available FaVAD tool will facilitate the computational research in the
area of ion-solid interactions. In particular, the provided probabilistic measure for atoms
being in a distorted environment and the easily comparable defect-fingerprints 
should foster a reproducible assessment of damage in MD simulations
in crystalline structures. In addition, FaVAD provides the full processing chain from 
the simulation output (atom positions) to the visualization.


\section*{Acknowledgments}
F.J.D.G gratefully acknowledges funding from the A. von Humboldt 
Foundation and the C. F. von Siemens Foundation for research fellowship.

\section*{References}
\bibliographystyle{elsarticle-num}
\bibliography{bibliography}

\end{document}